\documentclass[fleqn,twoside]{article}
\usepackage{espcrc2}

\usepackage{epsfig}
\usepackage[figuresright]{rotating}

\newcommand{\AmS}{{\protect\the\textfont2
  A\kern-.1667em\lower.5ex\hbox{M}\kern-.125emS}}

\title{NOSTOS experiment and new trends in rare event detection}

\author{ I.Giomataris\address[CEA]{CEA, Saclay, DAPNIA, Gif-sur-Yvette, Cedex,France.},S. Aune\addressmark[CEA],
P. Colas\addressmark[CEA], E. Ferrer\addressmark[CEA], 
I. Irastorza\addressmark[CEA], B.
Peyaud\addressmark[CEA], J. Dolbeau\address[PPC]{IN2P3/CNRS
PCC-Coll\`ege de France 75231 Paris Cedex 05 France}, P.
Gorodetzky\addressmark[PPC],
H, van der Graff\address{NIKHEF, Amsterdam The Netherlands},
 T. Patzak\addressmark[PPC], 
P. Salin\addressmark[PPC], J. Busto\address{CCPM-Facult\'e des
Sciences de Luminy, 13288 Marseille Cedex 09 France}, 
V. Lepeltier\address{LAL-B\^at. 200 F-91405 Orsay France},
V. Lepeltier\address{LBNL Berkeley USA},
 I. Vergados \address{University of Ioannina T.P.D., PO Box 1186, 45110 Ioannina
Greece}}
\begin{document}

\begin{abstract}
A novel low-energy neutrino-oscillation experiment NOSTOS, combining a strong tritium source and a high pressure spherical TPC detector (10 m in radius) has been recently proposed. The goal of the experiment is to measure the mixing angle $\theta_{13}$, the neutrino magnetic moment and the Weinberg angle at low energy. The same apparatus, filled with high pressure Xenon, exhibits a high sensitivity as a Super Nova neutrino detector with extra galactic sensitivity. Results of a first prototype will be shown and a short-term experimental program will be discussed.
\vspace{1pc}
\end{abstract}

\maketitle

\section{Introduction}

Nowadays there is a very strong evidence for neutrino oscillation from atmospheric and solar neutrino experiments. Recent results from the KamLAND confirm earlier work at the Sudbury Neutrino Observatory and Super-Kamiokande  that also provided strong evidence for neutrino oscillation. The fact that different neutrino types do change their identity as they propagate, suggests that they are not strictly massless as had been assumed by the standard weak interaction model.
The atmospheric neutrino oscillation data suggest that there is maximal mixing between the $\tau$ and e- neutrinos and a corresponding mass squared difference of $\delta m^2_{23}=3\times10^{-3} eV^2$ . On the other hand the solar neutrino and other electron neutrino disappearance experiments suggest a non maximal mixing of   and a mass squared difference of $\delta m^2_{21}=7\times10^{-5} eV^2$ . The other mixing angle   is not known but it is quite small constrained from the CHOOZ data. If one includes this small mixing angle, one can see in electron neutrino experiments the effect of the large $\delta m^2_{23}$, i.e. a transition probability associated with the small oscillation length. In other words for a detector close to the source one will see a disappearance oscillation probability of the form:
\begin{equation}\label{oscillation}
 P(\nu_e \rightarrow \nu_{\mu, \tau}) = \sin^2 2 \theta_{13}
 \sin^2\pi\frac{L}{L_{23}}
\end{equation}
where L is the distance between source and observed interaction, E
is the neutrino energy and L$_{23}$ is related to the neutrino
energy: $L_{23}=2\pi E_\nu / (\delta m^2_{23})$. In the case of
very low energy neutrinos, such as the ones emitted by a tritium
source, the oscillation length L$_{23}$ is only 13 m. Therefore by
observing neutrino interactions inside a large TPC of about 10 m
in radius, surrounding a tritium source, one can contain the
oscillation occurring inside the gas volume and measure the
oscillation parameters by a single experiment. The idea of
combining a strong tritium source and a spherical detector has
been recently proposed\cite{ref1}. In this proposal a large
spherical drift volume filled with a suitable gas mixture at high
pressures is used for detecting low energy recoils produced in
neutrino interactions.


\section{The spherical TPC concept}

The tritium source is located in a small spherical vessel of about
25 cm radius and it is surrounded by a spherical gaseous TPC 10 m
in radius as shown in figure 1. Low energy electron recoils
produced in the gas by elastic scattering in the TPC volume are
ionizing the gas. Charges are drifting towards the sphere center
and are collected by an adequate gaseous detector. The use of a
spherical Micromegas counter made out of flat detectors was
advertised because of the high precision and excellent energy
resolution\cite{ref2,ref3}. The high efficiency for detecting
single electrons has been demonstated \cite{ref4}even at high
pressures \cite{ref5}. The detector is currently used for solar
axion detection in the CAST experiment \cite{ref6}where  a great
stability and ability to reject background events has been
achieved.

The novel approach is radically different from all other neutrino
oscillation experiments in that the neutrino source and the
detector are located in the same vessel; it is then possible to
measure the neutrino interactions, as a function of the distance
source-interaction point, with an oscillation length that is fully
contained in the detector. Indeed we expect a counting rate
oscillating from the centre of the sphere towards the external
volume, i.e. at first a decrease, then a minimum and finally an
increase. In other words we will have a full observation of the
oscillation process as it has already been done in accelerator
experiments with neutral strange particles. Fitting such an
observed curve will provide all the relevant parameter of the
oscillation by a single experiment. It is equivalent to many
experiments made in the conventional way where the neutrino flux
is measured in a single space point. The use of a spherical TPC
detection scheme presents  many advantages :
\begin{itemize}
\item It is a natural focus device requiring a small amplifying detector with only a few read-out channels. Such small size detector simplifies the construction and reduces the cost of the project.
\item With the neutrino source at the center of curvature the detected signal is optimized with a gain of a significant factor compared to the signal provided by a cylindrical TPC with the detector aside the end-cup of the TPC.
\end{itemize}

The electric field at a distance r from the center of curvature,
in the case of a spherical TPC,  is given by :
\begin{equation}\label{electric_field}
E=\frac{V_0}{r^2}\frac{R_1 R_2}{R_2 - R_1}
\end{equation}
where R$_1$ and R$_2$ are the internal and outer sphere radii and
V$_0$ is the applied voltage. At low electric fields (E) the drift
velocity $v_D \approx E$ is roughly proportional to E and the
longitudinal diffusion coefficient is $D \approx 1/\sqrt{E}$,
inversely proportional to the square of E. From the previous
expressions one can easily deduce that time dispersion is :
\begin{equation}\label{diffusion}
 \sigma_t = \frac{\sigma_L}{v_d} = \frac{D\sqrt{r}}{v_d }\approx
\frac{1}{E^{3/2}} \approx r^3
\end{equation}
From the latter equation one can conclude that in the spherical detector there is enhancement of the time dispersion of the detected signal especially at large distances. We propose to use this property for estimating the depth of the produced electron recoil during the neutrino elastic scattering. First calculations show that a precision for the depth of the interaction point of better than 10 cm will be achieved by measuring the time dispersion of drifting charges.
Using a 20 Kg tritium source, the total rate of emitted neutrino is $6\times10^{18} /s$. With a gas filling of Xe at p=1 bar the rate of detected neutrino is about 1000/year. We are studying various issues for increasing the pressure in Xenon or other noble gases in order to get a larger signal or decrease the amount of the tritium source.

\section{Sensitivity for the neutrino magnetic moment}

Because of the low energy of the incoming neutrinos and the low energy electron recoils detected in this experiment the sensitivity for the neutrino magnetic moment is high. The cross section of the magnetic moment can be written as :
\begin{equation}\label{n_coherent}
\left(\frac{d\sigma}{dT}\right)_{EM} =  \sigma_0
\left(\frac{\mu_l}{10^{-12}\mu_B}\right)^2
\frac{1}{T}\left(1-\frac{T}{E_{\nu}}\right)
\end{equation}
Because of the dependance 1/T (T is the electron recoil energy)
the sensitivity for the magnetic moment is obviously higher at low
energy. Indeed precise calculations show that the differential
cross section due to a neutrino magnetic moment of 10$^{-12}$
$\mu_B$ is rising at low energy and reaches 30\%  of the value of
the weak neutrino-electron cross section which is at a first order
flat with the recoil energy. We would like to point out that
recent measurements from the NuMu experiment predict a limit of 10
$^{-10}$ $\mu_B$ for the neutrino magnetic moment. Our experiment
opens the way to improve this value by two orders of magnitude.

\section{The measurement of the weak charge}
Another interesting quantity is the Weinberg angle appearing in
$\sin^2 \theta$, which is   a function of the momentum transfer
and it has not been measured at such low transfers. To this end
atomic physics experiments, which utilize the neutral current,
have thus far been considered. The neutral current at low energies
gives a current-current interaction of the form:
\begin{equation}\label{weak charge}
 H_W = \frac{G_F}{\sqrt{2}}J_\lambda^Z J_Z^\lambda
\end{equation}
\begin{eqnarray}\label{weak charge2}
 J_\lambda^Z = \bar{p}\gamma_\lambda(1-4 \sin^2\theta_W
-\gamma_5)p+ \nonumber\\  + \bar{n}\gamma_\lambda(1-\gamma_5)n -
\\  - \bar{e}\gamma_\lambda(1-4\sin^2\theta_W-\gamma_5)e
...\nonumber
\end{eqnarray}
Thus the atomic physics experiments suffer  from the fact that the
weak charges involved are extremely small, i.e. the weak charge of
the electron is $-1+4\sin^2\theta =-0.1$, while for the proton is
0.1. Due to this smallness one has to deal with complications in
the analysis of the experiments arising from radiative
corrections. This has also implications in the neutrino nucleus
elastic scattering in the sense that the neutrons in the nucleus
can contribute coherently. The coherence due to the protons is
suppressed by the smallness of the weak charge.
 In the proposed experiment such cancellations do not occur and one needs not worry about such corrections. Furthermore by plotting the differential neutrino-electron cross section as a function of the electron energy we obtain a straight line. We hope to construct the straight line quite accurately. Thus we can extract a value of the Weinberg angle both from the slope and the intercept achieving high precision.

\section{Supernova sensitivity}

It is generally believed that the core-collapse supernova
explosion produces a large number of neutrinos and 99\% of  the
gravitational energy is transformed to neutrinos of all types. The
supernova (SN) neutrino flux consists of two main components: a
very short ($<$ 10 msec) pulse of $\nu_e$ produced in the process
of neutronization of the SN matter through the reaciont e + p
$\rightarrow $ e + n, which is followed by a longer ($<$ 10 sec)
pulse of thermally produced $\nu_e$, $\nu_\mu$, $\nu_\tau$, and
their antiparticles. Only a small fraction, about 1\%, of the
neutrinos are prompt, while the rest are neutrino pairs from later
cooling reactions. Therefore the neutrino signal from a supernova
rises first steeply and then falls exponentially with time in a
time window of about 10 seconds.  It is expected that spectra of
thermally produced neutrinos are characterized by the different
mean energies : $\nu_e=11$MeV, $\bar{\nu}_e=16$MeV,
$\nu_{e,\mu}=25$MeV. Our idea is to use the large cross section
offered by the coherent neutrino-nucleus cross section for
detecting neutrinos from Super Nova explosions. Coherent
scattering occurs when neutrinos interact with two or more
particles and the amplitudes from the various constituants of the
target adding up. A consequence of the coherence is an increase of
the cross section becoming proportional to the square of the
number of particles in the target leading to ncreased counting
rates  :
\begin{equation}\label{crosssection}
 \sigma = \frac{G^2N^2E^2}{4\pi}
\end{equation}
where G is the weak coupling constant, N is the number of neutrons
in the target nucleus and E is the neutrino energy. In order to
get advantage of the coherent scattering amplification of heavy
nuclei gases are needed.

For instance, using Xenon as detector target the coherent  cross
sections at  E=25  MeV, the energy that is relevant for Supernova
detection, is  quite  large ($\sigma =1.5 10^{-38}$ cm$^2$). Even
at lower energy  (11  MeV) where  the  coherent  cross  sections
decreases quadratically  with  energy,  the cross section is still
high, in the case of gaseous Xenon detector. The recoil energy
energy is quite low and it takes a maximum value of 1.5keV for 11
MeV and 9 keV for 25 MeV neutrinos. This implies that detector
thresholds must be set quite low with one advantage that
backgrounds are highly suppressed given the narrow time window in
which the burst takes place. The collected energy is even lower by
a significant factor (quenching factor) and therefore sub-keV
detector threshold is required as it is the case of the NOSTOS
spherical TPC. For a typical galactic SN explosion the detector
used for the tritium experiment (10 m in radius, p=10 bar of
Xenon) the number of detected neutrinos will exceed 100,000. Such
high signal will also permit extra galactic detection. A
possibility to test the efficiency of detecting coherent neutrino
scattering will be the nuclear reactor. The expected number of
neutrino interactions in the gas volume, from a typical reactor
neutrino flux and spectrum ($10^{13}$ cm$^{-2}$ s$^{-1}$) using a
detector filled with Xenon is about 350/day/Kg. The drawback is
the very low energy threshold needed since the maximal recoil
energy is 185 eV. Therefore single electron counting is required
imposing a high gain operation of the detector and a measurement
of the quenching factor of the ionization produced by the low
energy recoils. Such an interesting measurement could be possible
by using the first prototype that is described in the next
section.

\section{The 1st spherical prototype and results with a new proportional counter}
We have built a spherical prototype 1.3 m in diameter as a first
step towards the large detector needed for accomplishing the
NOSTOS project. A schematic view of the prototype is shown in
figure 2. The outer diameter is made of pure Cu (6 mm thick)
alowing to reach pressures as high as 5 bar. The 1 m3 gaseous
volume is pumped through a primary followed by a turbo molecular
pump. The quality of the various materials assures a good vacuum
($<10^{-6}$ mbar) with a quite low level of outgasing that has
been measured to be $<10^{-9}$ mbar/s.  Such low outgasing give us
the possibility to avoid a permanent gas circulation through
special cleaning filters. First tests were performed by filling
the volume with pure gas (argon + 10\% CO$_2$ $<$3ppm impurities)
and operate the detector in a seal mode. We are proposing the use
of Micromegas as amplifying structure. A particular effort is
actually made to build a spherical Micromegas detector using new
technologies. Another alternative will be to approximate a sphere
by using several Micromegas flat elements. Preparing an ideal
solution for the amplification structure we decided to start the
first tests by using a small sphere (10 mm in diameter made of
steel) as a proportional counter located at the center of
curvature of the TPC.  Several tests from low pressure to high
pressure have shown that such a simple amplification element is
able to provide high gains and stable operation. With outer sphere
at ground and the inner sphere at the High Voltage (HV), signals
produced by cosmic rays or a $^{109}$Cd radioactive source have
been observed at various pressures. A typical signal from the
cadmium source is shown in figure 3. It has a rise time of about 2
$\mu$s that is due to the movement of positive ions. This is quite
a long rise time but this simple structure will allow us to make
fast progress, understand the functioning of the spherical TPC and
accomplish part of the short term experimental program to
establish the new concept that can be summarized as follows:

\begin{itemize}
  \item Tests of the 1st prototype and optimization of the amplification
structure.
\item  Optimize the detector for very-high gain operation.
\item  Measure the attenuation length of drifting electrons.
\item  Optimize the energy resolution.
\item  Measure the accuracy of the depth measurement by the time dispersion of the signal.
\item  Optimize mechanics and electronics, use low-radioactivity materials.
\item  Improve the simulation program.
\item Calculate (or measure) the quenching factor in various gases (Xe, Ar..).
 \item Measure the coherent neutrino scattering next to a reactor.
\end{itemize}

\section{Conclusions}
An ambitious experimental program in the low energy neutrino physics sector is proposed. It includes observation of neutrino oscillations, neutrino magnetic moment and Weinberg angle at low energy measurement. The same device exhibits a high sensitivity super Nova neutrino detection with extragalactic sensitivity. The measurement of the coherent neutrino nucleus scattering will open the way to build simple, low cost and robust telescopes dedicated for super Nova neutrino detection.
A first prototype 1.3 m in diameter has been build giving first promising results. A novel concept of spherical proportional counter has been successfully tested.

\begin{figure}[htb]
\includegraphics[width=2.5in]{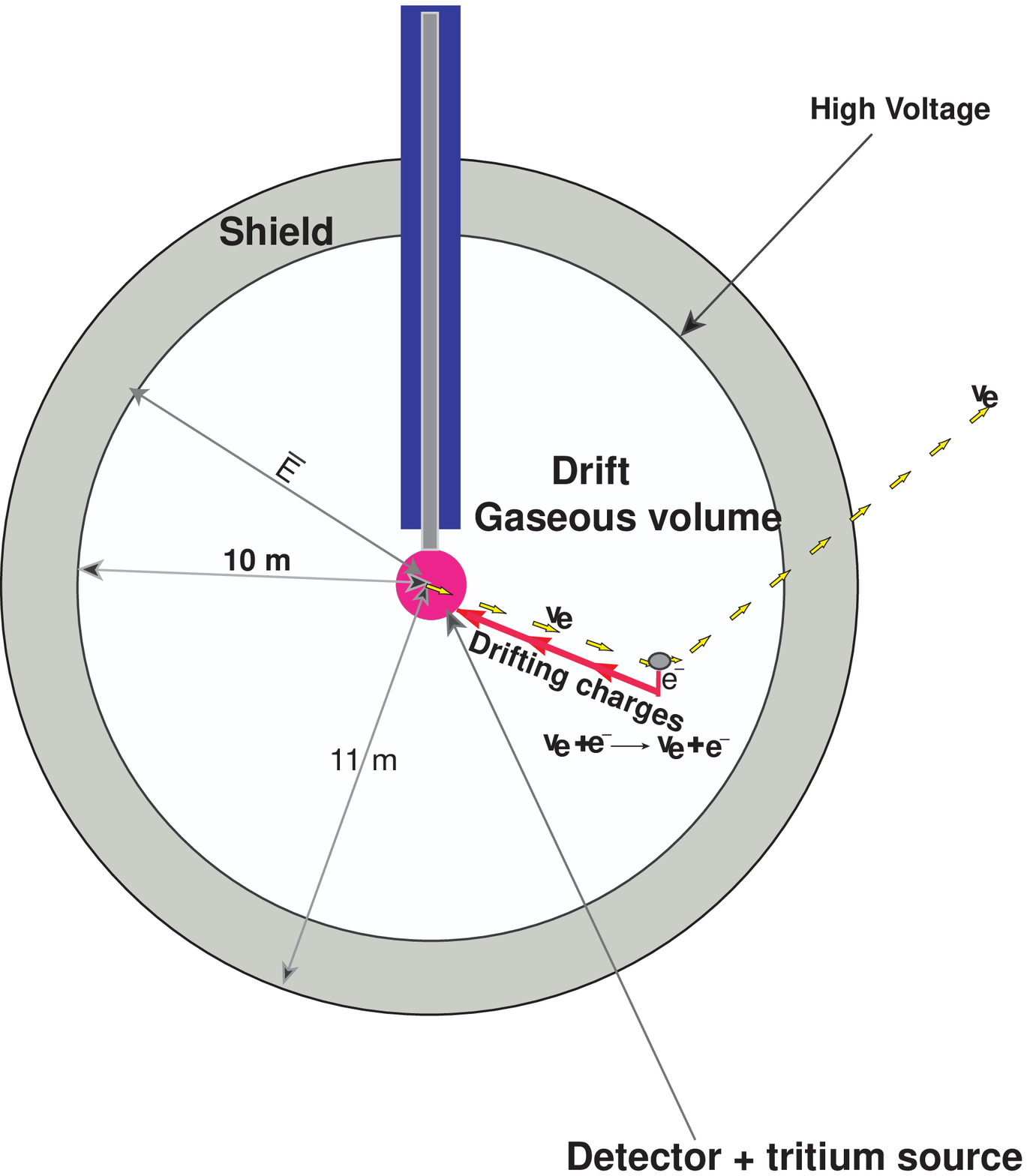}
\caption{Scheme of the NOSTOS detector} \label{fig:toosmall0}
\end{figure}

\begin{figure}[htb]
\includegraphics[width=2.5in]{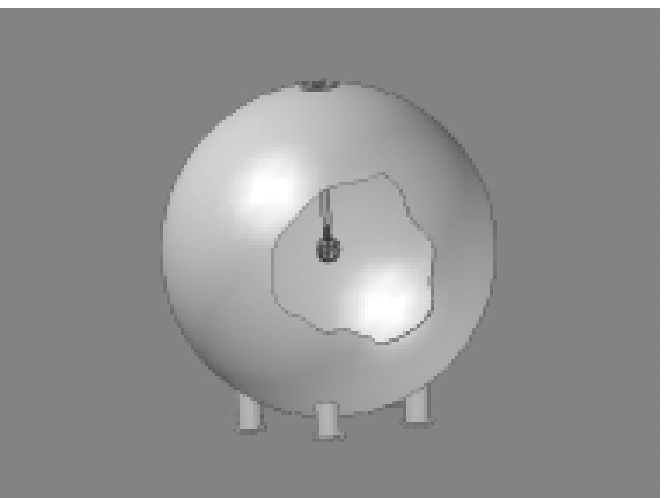}
\caption{Scheme of the NOSTOS detector}
\label{fig:toosmall}
\end{figure}

\begin{figure}[htb]
\includegraphics[width=2.5in]{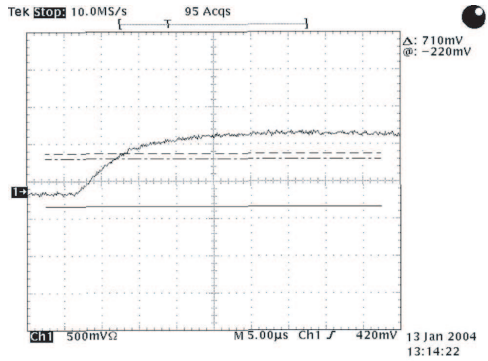}
\caption{A signal from the 109Cd radioactive source with the
detector filled with Ar+10\% CO$_2$ at P=600 mbar. The long tail
is due to the large RC of the charge amplifier.}
\label{fig:toosmall2}
\end{figure}

\begin{figure}[htb]
\includegraphics[width=3in]{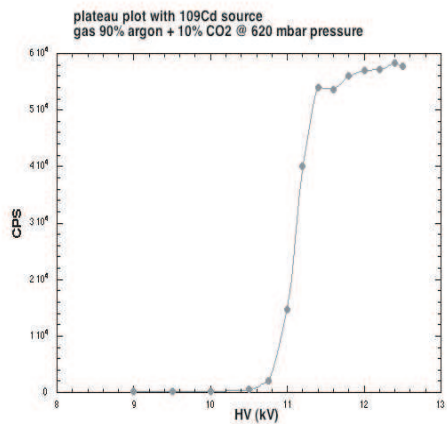}
\caption{Counting rate versus high voltage using a 109Cd
radioactive source.}
\label{fig:toosmall3}
\end{figure}

\end{document}